# Wave propagation in anisotropic crystal using point contact excitation and detection method

Varun Bhardwaj[1†], Kaushik Shukla[2], Frank Melandsø[3] and Anowarul Habib[3]
[1]Dept. of Chemistry, Indian Institute of Technology, Guwahati, Assam 781039, India
[2]Dept. of Elec. Eng., Indian Institute of Technology Dhanbad, 826004, India
[3]Dept. of Physics and Tech., UiT, The Arctic Uni. of Norway, 9037 Tromsø, Norway

## 1. Introduction

Lithium Niobate ($LiNbO_3$) is a piezoelectric crystal with a high electromechanical coupling coefficient. The development and miniaturization of acousto-electronics, and acousto-optics modulation filters, are primarily based on surface acoustic waves (SAWs) and bulk wave propagation in anisotropic crystals, such as $LiNbO_3$, Lithium Tantalate, and Quartz[1-3]. It has the potential to be used in nanoscale acoustics. The mechanical properties of the piezoelectric materials depend on the chemical compositions and crystal growth conditions. These excellent features are due to higher second-order optoacoustic nonlinearity with low losses in the broadband spectrum.

In the last several decades, the visualization of bulk acoustic and surface acoustic waves (SAWs) in the anisotropic piezoelectric crystal has been an active area of research. The generation and detection of bulk and SAWs in piezoelectric crystals with the help of an interdigital transducer (IDT) have attracted widespread scientific interest for signal processing and filtering applications[4]. The directional-dependent acoustic wave velocity from the spatial-temporal information of wave propagation depends on the characteristics of the crystal. The propagation of bulk waves in the piezoelectric crystals causes the transformation to dispersive Lamb waves. Lamb waves are dispersive guided waves that travel in two orthogonal wave modes called symmetric and anti-symmetric wave modes This paper proposes a point contact excitation and detection scheme to visualize metamorphosis of bulk wave to Lamb wave. The presented technique visualizes the amplitude of the acoustic waves and transforms the bulk waves into Lamb waves with sub-micron resolution. This technique utilizes the transfer of electromagnetic field to mechanical energy to excite phonon vibration in piezoelectric materials. The gradient of electric field and piezoelectric properties control the electromagnetic coupling.

------------------------------------------------------------

email: anowarul.habib@uit.no

## 2. Experimental Setup

For the last several years, our group has dedicated efforts to optimizing the point contact excitation and detection method for generating and visualizing broadband ultrasonic waves in piezoelectric materials [5-9]. Our group has previously provided a comprehensive overview of the excitation and detection probe fabrication, and experimental setup. The piezoelectric coupling property of the anisotropic crystal enables the steel probe to convert electromagnetic waves into mechanical waves.

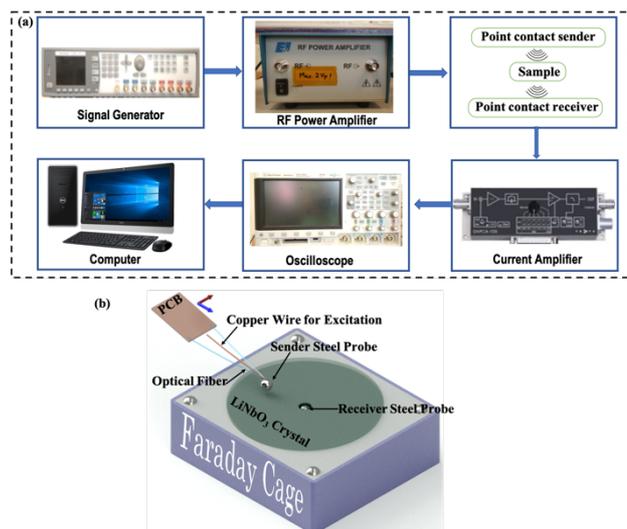

**Fig. 1:** a) Schematic representation of the point contact excitation and detection method, and b) detailed overview of excitation and detection setup.

The excitation probe in the current experimental setup emitted ultrasonic waves at each pixel, while the detector was stationary and positioned in the middle of the scan area. A 35 ns pulse excites the acoustic wave in a $LiNbO_3$ (X-cut, 1 mm thick, optically polished on both sides) crystal. A trans-impedance amplifier boosts the received signal. The time-resolved voltage data at each pixel location provided two-dimensional spatiotemporal imaging of acoustic wave propagation.

## 3. Result and discussions

3-D finite element method simulation of ultrasonic waves in LiNbO$_3$ crystal in the time domain had conducted using COMSOL Multiphysics 5.6 version. 4 mm x 4 mm X-cut LiNbO$_3$ of thickness 1 mm is the geometry used for simulation.

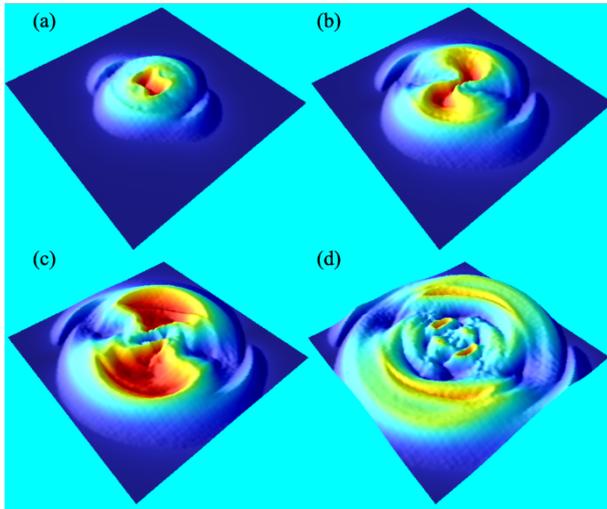

**Fig. 2:** Pseudo 3D representation of the ultrasonic wave propagation on X-cut 1 mm thick LiNbO$_3$ crystal.

Piezoelectricity Multiphysics in the Electromagnetics - Structure interaction subsection of the Structural Mechanics section is used for simulation. A low reflecting boundary condition is applied to all the faces except the upper flat surface to absorb all the outgoing waves to prevent unwanted reflections and interference. Each frame of fig. 2 corresponds to a 50 ns time interval in the transient signals. Under Electrostatics physics, the electrostatic potential and ground boundary conditions are added to the top and bottom surfaces. An input excitation signal uses a rectangular excitation pulse with a width of 20 ns, and an input voltage of 10 V. Time-dependent analysis and a time-dependent solver are utilized in the simulation. We have demonstrated the visualization of wave propagation in LiNbO$_3$ crystal using point contact excitation and detection method. The scanning area was 20×20 mm$^2$. After the acquisition of transient signals for every position of the scanner, time gating was performed to construct the snapshot of the ultrasound waves. Fig. 3 represents the pseudo-3-D evolution of ultrasonic waves in LiNbO$_3$ crystal at an excitation gate width of 30 ns and the time difference of each image is 25 µs. As wavelength increases and is comparable to the thickness of the LiNbO3 crystal, it interacts with the edges of the sample. Due to multiple reflections of the waves with the traction-free surfaces of the crystal, bulk wave metamorphosis into Lamb waves occurs.

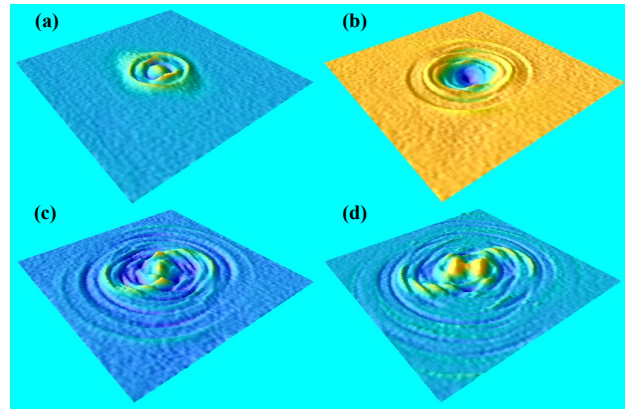

**Fig. 3**: Pseudo-3D represents the spatial and temporal evolution of the acoustical amplitude of images on X orientation 1 mm thick, LiNbO$_3$ crystal. The sequential images (left to right) show the evolution of transversal bulk waves and their metamorphosis into Lamb waves.

## 4. Conclusion

Point contact excitation and detection method have been developed for visualizing the evolution of bulk waves (high frequency-low wavelength) and their metamorphosis to Lamb waves due to multiple reflections in LiNbO3 crystal. Narrow pulse width and wide Fourier spectrum ensure metamorphosis of transversal bulk waves into Lamb waves for scan lengths comparable to the involved wavelengths and characteristic thickness.

**Acknowledgment**
The authors acknowledge the funding from the Research Council of Norway, Cristin Project, ID: 2061348.